# The Security of WebRTC

Ben Feher, Lior Sidi, Asaf Shabtai, Rami Puzis
Ben-Gurion University of the Negev
{feherb, liorsid, shabtaia, puzis}@bgu.ac.il

## ABSTRACT
WebRTC is an API that allows users to share streaming information, whether it is text, sound, video or files. It is supported by all major browsers and has a flexible underlying infrastructure. In this study we review current WebRTC structure and security in the contexts of communication disruption, modification and eavesdropping. In addition, we examine WebRTC security in a few representative scenarios, setting up and simulating real WebRTC environments and attacks.

## 1. INTRODUCTION

Web Real-Time Communication (WebRTC) is an API aimed to support browser-to-browser communication. This solution is designed to unify the fragmented solution cluster that has been dominating the market thus far. All innovations so far are standalone programs or browser plugins. These restrictions limit these solutions' integration into a growing browser-based world.

WebRTC comes as a solution for P2P file sharing, streaming P2P video and audio calls and soliciting real-time communication solutions into the end user's Web browser.

Since it was first presented, WebRTC has been considered a disruptive force demonstrated by the amount and quality of services currently using and underlying the WebRTC solution in their products. Some representative examples are Facebook's[1] use of this technology in its mobile device messenger application, Firefox Hello,[2] as well as addLive,[3] which has been acquisitioned by SnapChat. The technology's effectiveness is especially visible by the number of startup companies that grew from it, demonstrating that this tool is not restricted just to the enterprise level such as VOIP, and can be leveraged by anyone.

In this study we review WebRTC security, while focusing on the main players in a WebRTC application, exploring their security measures, such as access control, credential storage and cryptography. In addition, we analyze the WebRTC technology and present potential threats to a real world WebRTC featuring application [1].

As a multi-platform technology, different implementations of WebRTC may obey or disown RFC specification requirements. In addition, each WebRTC implementation may differ in uses (voice, video, chats, or online augmented reality). This implies that security evaluations must be performed for each implementation specifically. As WebRTC client applications are, at the moment, primarily browser–based (though not all browsers support WebRTC) we focus our analysis on WebRTC browser-based impelmetations. A quick look at browser usage statistics reveals that currently, Chrome and Firefox dominate the browser market by a significant margin. In addition, these browsers also support key features of WebRTC technology which makes them prime candidates for our tests.

Another point for consideration in the security analysis involves the communication restrictions of the current security measure (e.g., deep packet inspection Firewalls), as well as the availability of existing tools for analyzing the WebRTC traffic.

## 2. WHAT IS WebRTC?

In General, a WebRTC communication is composed of two stages: a signaling and a communication stage.

### 2.1 Signaling Stage

The signaling stage is the conversation setup. When two clients wish to communicate, they must arrange a common information channel. If client *A* wishes to communicate with client *B*, Client *A* must convey an address at which Client *B* may reach him. In telephony, this stage is attributed to Client *A* calling Client *B*. Client *A* dials client *B's* phone number, and starts the call. The telecommunication company is now responsible of finding *B* in its network and notifying him that *A* is calling him. In turn, *B* can choose to answer, reject or ignore the call. All these options must be relayed back to Person *A*. In this scenario, the phone number is the address at which to reach client *B*.

The general (common) WebRTC communication scheme is presented in Figure 1. When client *A* wishes to communicate with client *B*, the steps performed are:

(1) Client *A* opens a local listening connection and sends a request to the signaling server to communicate with *B*. This request hosts information about the client's capabilities (available media codecs, API version, etc.), as well as unique security identifiers: (a) unique session ID, (b) session start/end time, (c) session version (d) media stream identification, (e) ice-ufrag/ice-pwd – these values are used to initiate direct connection between peers. Once the direct P2P connection is established, these values will be used to authenticate each connection to the other, (f) DTLS fingerprint – when the two clients open a direct connection between themselves, they will use an asymmetric encryption for their communication, similar to TLS. When the two parties commence with direct communication, each client sends the other a TLS certificate; this fingerprint is used to ensure the TLS certificate's authenticity once received.

(2) The signaling server (optionally) authenticates *A,* and forwards the request to *B*.

(3) Client *B* approves/rejects client *A*'s request. If the request is approved, Client *B* opens a listening connection and sends the response to the signaling server.

(4) Signaling server forwards response to Client *A*.

---
[1] https://webrtchacks.com/facebook-webrtc/
[2] https://support.mozilla.org/en-US/kb/firefox-hello-video-and-voice-conversations-online
[3] https://bloggeek.me/snapchat-acquires-addlive/

(5) Assuming client *B* approved the request, client *A* and *B* exchange direct communication information. This includes IP where a listening interface is open for communication, connection type (TCP/UDP), and possibly NAT traversal information (IP/port bindings).

(6) Both clients open direct communication channel with each other.

A pre-condition for establishing WebRTC communication between *A* and *B* is that both must use an application or a Web service which implements/supports the WebRTC communication. This may be set using a browser where the clients browse to an online address or a locally stored page, or using a native application (e.g., Java, C++) with pre-stored local and remote configurations (Stage 0 in Figure 1).

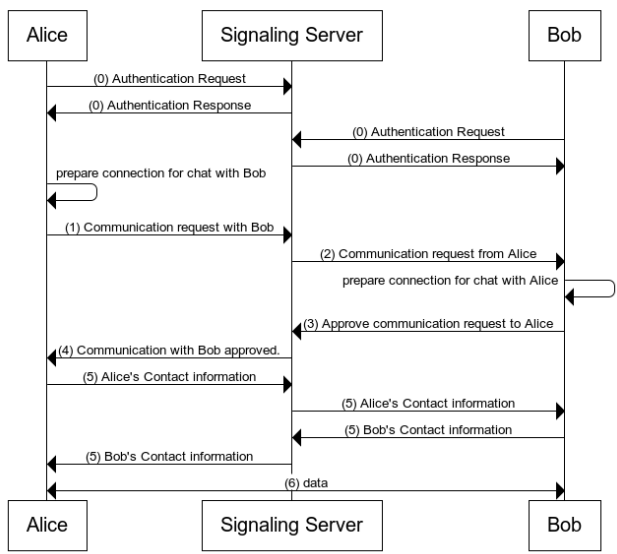

**Figure 1**: Example Of conversation process.

## 2.2 Communication Stage

Once participants have shared their information with each other, a direct channel is established and from this point on the two participants can exchange information without the signaling server's intervention. Since this is a new connection, both clients must re-validate each other. This is performed by re-sending information previously sent in the signaling phase (ice ufrag, ice pwd and DTLS certificate). Once both clients validate each other, they may communicate directly with each other. This stage differs from classic web communication due to the fact that communication is now server-less. In all classic browser communication platforms, each message sent from user *A* to user *B* is sent through the server. This increases latency and is a major breach in the user's privacy.

## 2.3 WebRTC Gateway Typology

WebRTC Gateways are mainly designed to allow telecommunication providers to harness WebRTC technology, while maintaining critical functionalities such as billing and lawful interception. The main difference is adding a proxy server. If Caller *A* wants to call recipient *B*, he actually performs a call to a proxy server *C*. Figure 2 illustrates the signaling process: (1) Alice sends a communication request to Bob's virtual client via the signaling server.

(2) Alice's Virtual client initiates communication with Bob via the signaling server.
(3) Bob approves communication with Alice's virtual client.
(4) Bob's virtual client approves communication with Alice.
(5) Both virtual clients initiate communication with each other.
(6) All communication between Bob and Alice are tunneled through their virtual clients.

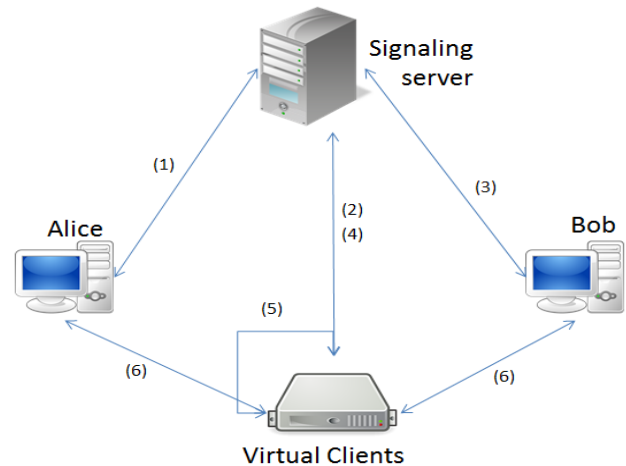

**Figure 2:** Illustration of the second WebRTC topology model.

## 3. SECURITY IN WebRTC

WebRTC security measures can be divided into three main groups: Client, Server, and Network Traffic.

## 3.1 Server Security

Since WebRTC is designed as a Peer-to-Peer communication interface, the server's duty is the mediation of clients. As such, the roles performed by the server are implemented using pre-WebRTC software components, such as STURN (Session Traversal Utilities for NAT) and IdP (Identity Provider). No additional designated security measures are created.

## 3.2 Client – Access Control

Since WebRTC is primarily designed to facilitate P2P audio and video chats, the facilitating browser should be able to access the microphone and video camera. This requires a user's consent. Major browsers implement two mechanisms. Media devices' permission requests should be requested per use, unless stated otherwise by the user. Users must be informed of a media device being used in a clear and visible manner at all times of the resource's use [2]. A few examples of the access control mechanism are presented in Figure 3.

## 3.3 Network Traffic – Channel Encryption

### 3.3.1 Encryption

WebRTC specification explicitly forbids clear RTP (Real-Time Transport Protocol) [3] and enforces using a secure encrypted version of it called SRTP (Secure Real-Time Transport Protocol). For this to be achieved, we must make the distinction between the signaling plane (HTTP, SIP) and the media plane (RTP). Each of these can be secured independently but this may be troublesome. This mechanism does not guarantee that the signaling user is the same as the media messaging user. Due to this deficiency, WebRTC specifications enforce encryption of both channels using DTLS-SRTP. This ensures that the establishment of the media stream can take place without the need to reveal the Session Descriptor Protocol in the message exchange.

The DTLS-SRTP protocol is a variation of TLS, commonly used in an HTTP based session to achieve secure sessions between a client and a server (HTTPS). It is modified to accommodate the needs of live human to human communications, that is for example, when sending a file, the packet order is not important as data can be ordered upon data transfer completion [2].

Since TLS operations are considered to be expensive, SRTP (Secure Real-time Transport Protocol) is used as a lightweight encryption model for communications. DTLS-SRTP utilizes DTLS to bootstrap the SRTP key exchange over a higher encryption level channel to prevent MiTM attacks [1].

Another point to consider in this aspect is how the browser reacts to a proxy that tempers with encrypted certificates. In normal HTTPS applications, there is a clear warning in each browser that the channel is compromised. This warning allows a client to make an informed decision about using a non-secure connection to the web application. Under WebRTC this choice is not given, and such conversations are rejected by the browser.

### 3.3.2 Certificate exchange

When listening on a line between parties of a conversation, an attacker may want to be able to listen in to the communicating parties by performing a man-in-the-middle (MiTM) attack.

Much like in standard TLS, certificate exchange is performed between the two parties. But two clients are not part of any Certificate Authority (CA), and a self-signed certificate is sent between them. In order for B to be able to guarantee A's certificate is authentic, the TLS certificate is sent the first time via the signaling channel, and then sent again via the media channel. At this point, the two certificates are compared for validity [4].This approach mitigates a MiTM attack as forging the certificate on the signaling channel requires breaking the SSL encryption to the signaling server.

### 3.3.3 Client denial of service

An attacker may attempt to prevent a client from being able to use a WebRTC application by overloading the client's resources (bandwidth, processor or screen). In the proposed WebRTC architecture, finding a client requires using the identification process and respectful component, which allows the system administrator to monitor any suspicious activity and mitigate it thus. In addition, the dynamic architecture of the Internet allows a user to switch IP addresses in the case that a previous SDP is used.

## 4. THREATS IN WebRTC

At this point, we chose a "hands on" approach to exploring the existing threats on a WebRTC application. An attacker may be trying to obtain information and masquerade as a participant, along with arbitrary activation of media devices.

We first decided to look at all the players involved: client, server, and information channel—and for each player, we searched for probable threats and their effects on the system:

Client – What are the effects of JavaScript/HTML injection?

Client – Can we steal WebRTC credentials?

Client – Can we steal privileged information about a client?

Server – What are the effects of taking over a signaling server?

Server – Can we crash the server, or render it unresponsive?

Information channel – What information can be extracted?

Information channel – Can we cause a client to connect to a rouge network?

For each player, we researched whether it is possible to disrupt a conversation, steal information or impersonate a user.

The evaluation and analysis of the threats are summarized in Table 1.

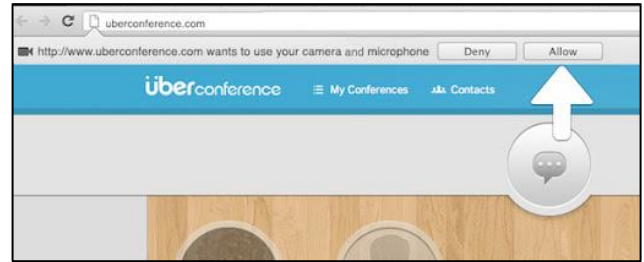

(3a) User is prompted for media device permission.

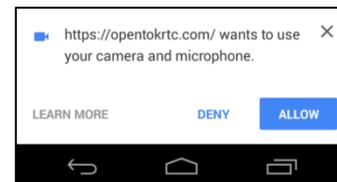

(3b) User is prompted for media device permission (Android).

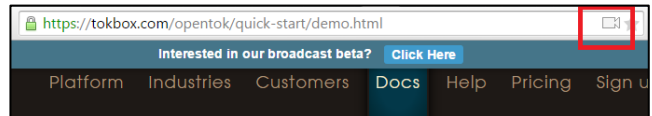

(3c) User is informed while media device is in use.

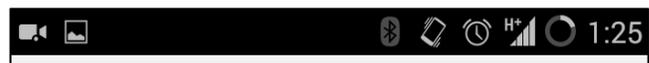

(3d) User is informed while media device is in use (Android).

**Figure 3**: Examples of access control, when a client is using the camera; an icon indicating this will appear in the browser in a clear manner.

**Table 1:** Summary of threats.

|  | Server | Client (Browser) | Server | Traffic | Client (Android) |
|---|---|---|---|---|---|
| **Threat** | Malformed JSON | JavaScript injection | Takeover by code execution | In line eavesdropping | Information exposure via malware |
| **Description** | An attacker sends malformed JSON to server as a part of a legitimate request | Inserting JavaScript into client's view of the website | Attacker fond a weakness in the code that allows code execution | Attacker taps the communication line to access residual data of WebRTC communication | Malware attempts to disclosure information |
| **Possible outcome** | Server crashes | Information exposure, identity theft | Information exposure, identity theft, denial of service | Information exposure | Information exposure, identity theft |
| **Mitigation** | Input validation, server auto reboot. | Input validation in server and client | - | - | Applying Android guidelines, certificate pinning. |

### 4.1 Client - Browser

While WebRTC Implements security measures to the extent of its context, it is important to note that WebRTC lives within a hosting browser. As such, a conventional web application attack effects the WebRTC environment and can be leveraged to gain information and assume actions on a victim's behalf.

### 4.1.1 JavaScript injection

**Description**: In this section, we assumed that an application is vulnerable to cross site scripting, allowing a user to insert HTML and JavaScript in the context of the application. We tested related security threats due to this flaw.

While classical XSS (cross-site scripting) attacks are mostly mitigated by input validation at server side, WebRTC applications allow users to share information directly without a server as a mediator. This relation means that an attacker may send malicious code to a victim embedded within a text message, a display name or an attachment file name, that is sent in a P2P fashion. Mitigating this requires the receiving end to perform input validations on received or displied user supplied data. Malicious content may be executed in specific scenarios but not others. For example, the malicious payload may be filtered in the chat pane, it may be executed in the chat history pane [5].

**Possible outcome**: Attacker may be able to perform some or all the following: extract a password using phishing and/or send messages on the WebRTC platform of the site and capture conversation data. This vulnerability has been shown in multiple occasions, including our own research which leverages any part of a WebRTC application that is exposed to an HTML/JavaScript Element [6]. An example of such an attack is showin in Figure 4.

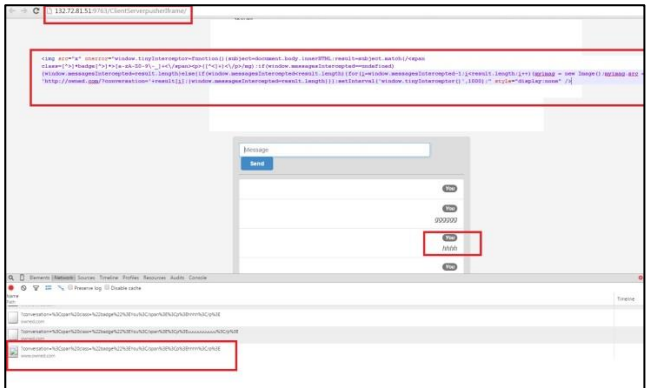

**Figure 4**: WebRTC chat is susceptible to an XSS attack, this allows chat user *A* to perform actions on the behalf of user *B*. in the same manner, an attacker can leverage a persistent or non-persistent XSS to sniff information, temper with chat participants and even gain privileged access to a server.

### 4.1.2 Leveraging signaling for information disclosure

#### 4.1.2.1 Local client

**Description:** Due to the nature of WebRTC, the browser must be able to access relevant information, such as the internal IP address of the machine. In a WebRTC application flow, this information will be used to establish a P2P connection without a server. Since this is a JavaScript API, it can be used in the context of any web application, allowing a server to collect this information from any client that uses its services in a browsing context. This problem was addressed from the construction of WebRTC [7] [8], but currently is still unsolved. A proof of concept of this was performed and is shown in Figure 6. Exploitation of this method requires the victim to have WebRTC enabled on his browser, once the victim browses to a malicious website, the attacker can activate WebRTC abilities via JavaScript code in order to gain sensitive information.

**Possible Outcome:** This allows, in principle (and practice), an attacker to gain useful information about the victim and his or her network [3].

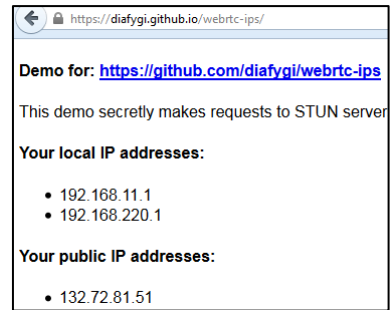

**Figure 5:** Demonstration of extracting an internal IP address via WebRTC.

#### 4.1.2.2 Remote Client

**Description:** As described in the introduction, P2P communication via WebRTC requires both clients to have a communication address of the other peer. This communication address reveals information such as the client's internal IP address. This knowledge allows an attacker to gain information about clients by opening a successful communication connection to them. Mitigation of this problem can be performed by applying an application architecture where a virtual client is a proxy to all the communication (as shown in section 2.3). By applying this architecture, the attacker can only view the virtual client's IP address, leaving the victim's IP known only to the application provider.

**Possible outcome:** Gain privileged information, for example, an attacker may be able to map teams and roles of personal within an organization by mapping their internal IP address. This attack relays on the fact that members of the same team are usually positioned in the same location, therefore, will likely share the same subnet [8].

### 4.1.3 Malware facilitation using WebRTC

**Description:** Web applications that allow users to share files amongst each other perform a security check on the file before sending it to the receiving party. A simple example of this is an email service blocking .EXE files to be sent. This test must be done at the server side as an attacker can edit a legitimate request after it was approved by the client side. In a WebRTC application, a conversation may take place where one client sends malware to another without any server intervention or knowledge.

Since files are not delivered via a server, it is important to define a location where the file's contents may be sanitized (searching for viruses) and validated (digital signature, matching file type, etc.) prior to its deliverance to the receiving clients. Three models exist for such a case:

a) (1) Client *A* sends file to server. (2) Server checks the file, (3) server sends file to client *B*.
 **Advantages:** Proprietary code for validation and sanitation remains private in server.
 **Disadvantages:** High bandwidth cost, each file is sent from and to the server.
b) (1) Client *A* sends file to server. (2) Server checks the file, (3) server sends hash of the file to client *B*. (4) client *A* sends Client *B* the file. (4) Client *B* computes file's hash against the server's sent hash.
 **Advantages:** Proprietary code for validation and sanitation remains private in server.

**Disadvantages:** High bandwidth cost, each file is sent to the server. Two files may be created to produce the same hash, one to be sent to the server and one to be sent to the receiving client [9].

c) (1) Client *A* sends client *B* the file, (2) client *B* checks the file.
   **Advantages:** Low bandwidth costs, all bandwidth costs are payed by the clients.
   **Disadvantages:** Private IP (Intellectual Property) of file sanitation and validation process will be shared with the users.

**Possible Outcome:** An attacker may send a victim a malicious payload to be executed by the victim's operating system.

## 4.2 Client – Registration and Termination

In this segment, we refer to considerations that must be taken for a WebRTC application in the context of registration and termination. These considerations are especially critical for an application that uses WebRTC as a replacement for GSM (Global System for Mobile) communications, with an emphasis on an application that allows a client to be connected in multiple locations at the same time (e.g., Apple Handoff [10]). Correctly registering and identifying users is critical to such services for monetary (billing) and security (lawful interception) reasons.

### 4.2.1 Registration
A client may register to a WebRTC application while providing no identifying details or allowing masquerading as a legitimate client.

4.2.1.1 Registration authentication via email

**Description:** An attacker opens an email account at some domain using no identifying features, then, proceeds to register to the WebRTC application.

**Possible outcome:** an attacker can use the application, allowing him to provide false registration details.

4.2.1.2 Registration authentication via phone number

**Description:** Registration authentication may be bypassed using one of the following scenarios:

(1) An attacker may use an Internet service for receiving SMS's that requires no identification to register to the WebRTC application.
(2) An attacker may use a pre-paid phone that requires no identification (some countries prohibit this).
(3) An attacker uses a MiTM attack to access the verification SMS sent to a GSM client in a proximity allowing GSM data interception. This attack, while complex, was shown to be possible [11], [12].

**Possible outcome:** an attacker can use the application, allowing him to provide false registration details.

### 4.2.2 Session termination
**Description:** Assuming an application allows a client to be connected in several locations at the same time (i.e., laptop and cell-phone), appropriate session termination mechanisms must be implemented. An attacker may be able to reuse an active session in a previously used (still logged in) device (e.g., a coffee place computer).

As a countermeasure sessions must be terminated once the system decides a device is compromised (e.g., by using anomaly detection solutions). One example of this is detecting when two devices are active under the same user in two remote geographic locations). Furthermore, the system must inform and enforce the device's session termination from existing active conversations. Since WebRTC allows for P2P actions, some actions may be performed without the server's intervention. This means that an attacker may perform actions under a victim's behalf (such continuing a conversation) while the victim's session is terminated in the compromised device.

**Possible outcome:** the attacker can steal private information and execute actions under a client's identity.

## 4.3 Client - Android
WebRTC can be implemented in Android in two ways: browser and native application. Native applications may embed an internal browser. In Android releases up to 4.3, WebView (browser API for native applications) does not support WebRTC capabilities. On the other hand, frameworks such as PhoneGap (Mozilla Cordova) allow WebRTC communication in a cross platform Web-Mobile Hybrid application. As a side note, there is always a possibility to take the piece of WebRTC code from an open source browser such as Chromium or Firefox, and integrate it into a native application. In this context, root privileges take an important role, as some features can be accessed only with root privileges.

### 4.3.1 Installation permissions
**Description:** An attacker creates an application that utilizes WebRTC to steal video/audio information.

**Possible outcome:** An attacker may leverage this application to capture private information about the user.

**Android Solution:** Installation of an Android WebRTC application prompts the user for specific privileges. At the minimum, a WebRTC application requires INTERNET permission. Telecommunication features also require RECORD_AUDIO and CAMERA permissions.

### 4.3.2 Credentials storage
**Description and possible outcome:** A malicious application tries to the capture credentials of an existing/installed WebRTC-based application.

**Android Solution:** By default, Android applications store application data in the /data/data folder (i.e., the applications' private folder). In order to read information from /data/data or any of its subfolders, root privileges are required. This means that unless the developer strictly defined otherwise, access without root privileges requires leveraging a flaw in the Android OS.

### 4.3.3 Tempering with the network traffic
We attempted to steal information from WebRTC applications in different settings: browser plugin, proxy, Android VPN (Virtual Private Network). While the conversation is protected by DTLS-SRTP, a client's credentials may be sent using an unencrypted channel beforehand, allowing an attacker to re-use the session. This scenario may take place when session credentials are sent over an unprotected channel, i.e., a non HTTPS connection to the WebRTC application server. In such a case, the attacker may masquerade as the legitimate client as well as acquire confidential information.

4.3.3.1 Browser Plugin

**Description:** An attacker attempts to install a browser plugin in order to steal WebRTC application credentials.

**Possible outcome:** The attacker assumes the victim's identity.

**Android Solution**: Testing was done on the two major browsers supporting WebRTC: Chrome and Firefox. While the mobile version of Chrome does not allow plugins, Firefox allows plugins the same abilities as it does to desktop applications. Browser plugins enable enhanced abilities to the browser, and have elevated permissions to define, capture and modify browser configurations—with defining a proxy for the browser being one of them. It is important to note that changing these configurations is not possible silently, and requires user approval.

4.3.3.2 Android Proxy

**Description**: A malicious application tries to redirect traffic to an unwanted destination by creating a proxy.

**Outcome**: The attacker gains sensitive information.

**Android solution**: While proxies can be defined by a user, in order for an application to configure a proxy on an Android phone, root permissions are required.

4.3.3.3 Android Proxy via VPN

**Description:** A malicious application tries to redirect traffic to an unwanted destination using a VPN Proxy.

**Outcome**: The attacker gains sensitive information.

**Android solution**: The Android VPN was investigated as an alternative to defining a proxy as the majority of applications that limit network traffic, as well as proxy applications such as Drony, Sendro and more, utilize this functionality. Defining a private VPN does not require a root, but requires a user's physical (click/touch screen approval) permission and cannot be performed silently.

### 4.3.4 Credential theft via UI Phishing

**Description:** (1) an attacker may create an application that performs the following: (a) the application UI impersonates the UI of an application to be attacked. (b) The application runs a service that checks what application is currently running in the foreground. (2) The attacker manages to install the malicious application to the victim's android device. (3) The malicious application runs in the background waiting for the application it is meant to impersonate to be opened. (4) Once the application is opened, the service opens the malicious application's UI, prompting the user for credentials. (5) The malicious application sends the information to the attacker and closes itself, returning the victim to the legitimate application [13].

**Possible outcome:** Credential theft, leading to information theft and actions performed in the victim's behalf.

### 4.3.5 Application Impersonation

**Description:** An attacker creates a clone of a legitimate application, and adds malicious code to the cloned application. This malicious application is then circulated to victims via social engineering.

**Possible Outcome:** private information theft, including credentials, leading to actions performed under the victim's behalf in the application's context.

## 4.4 Server

### 4.4.1 Signaling Server Takeover

Assuming an attacker can gain access to the signaling server, it is possible to:

(1) Crash the system.
(2) Connect random users into a conversation with each other.
(3) Forward communication by creating an "invisible" user participating in a conference conversation; in Figure 6 we have shown an example of such an attack.
(4) Perform a MITM by creating a fake user; each client will communicate with the fake user, believing they are communicating with each other. The fake user will proxy the conversation.

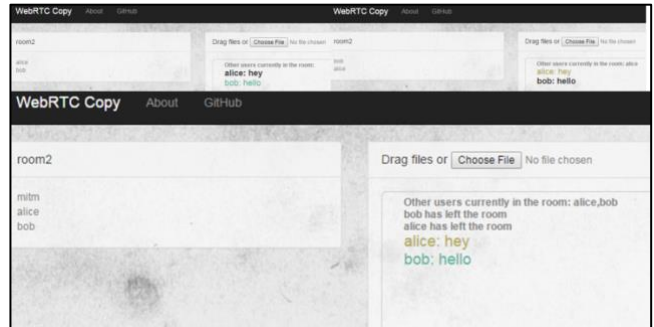

**Figure 6:** [Top left to bottom right] Victim A, victim B, attacker. This demonstrates a MiTM attack in a WebRTC chat where the server is compromised.

### 4.4.2 Crashing the server using malformed JSON

In this section, we have explored the ability to attack a WebRTC signaling server by sending a malformed JSON to the signaling server. The JSON object is parsed as one unit, meaning that a fault in one part of the object, will cause an error to the whole parsing process.

**Description:** An attacker sends a malformed JSON to the server as a part of a legitimate request (i.e. a new field in the ICE candidate message). When the server attempts to parse the JSON, it will produce an error and exit its main program loop.

**Possible outcome**: Server exits main program loop due to an error in parsing the malicious JSON. The aftermath of this is that the server is now unresponsive to new requests.

### 4.4.3 Registration Hijacking in SIP

**Description:** Registration to the Idp can be provided by several protocols, as long as the protocol fulfills the WebRTC contracts. A common communication signaling protocol used heavily in VoIP applications is SIP (Session Initiation Protocol). In this protocol, a client may register itself to the server, sending information such as an IP address where they can be contacted at. This information is sent in plain text. As it is being sent in plain text, an attacker may capture and modify it so all calls will be forwarded to the attacker [1].

**Possible Outcome**: An attacker may masquerade as a different user, receive and perform actions at the victim's behalf.

## 4.5 Information Channel

In this section, we detail how we tried to extract useful information, disrupt communication, or modify information by standing anywhere between the two clients.

### 4.5.1 Concluding from encryption

While WebRTC does hold a standard encryption to its information, it may still be possible to obtain some insight about the conversation type of two participating users and perhaps even extracting other useful information without tempering with the encryption. Mauro and Longoa (2015) showed a very high percentage of WebRTC communication classification, which brings about the question of whether an attacker can classify sub categories of WebRTC communication (e.g., a file is sent, audio call, video call), as well as additional useful information (if we are able to approximate the size and/or determine the type of a file that was sent).

### 4.5.2 Communication Disruption/DOS

Johnston, Yoakum, and Singh (2013) mentioned that enterprise firewalls allow for WebRTC communication due to false identification of a message as a response to a request. This suggests that an attacker can send packets to a client, fooling the firewall into thinking that these packets are a legitimate part of the conversation. Due to the UDP nature of WebRTC, IP spoofing is possible. At this point it might be possible to cause the firewall to block the legitimate conversation, though more research is still required [14].

### 4.5.3 Forceful use of rouge WI-FI network by leveraging server anti-DOS mechanism

**Description:** One way of protecting servers from DOS (Denial of Service) attacks is to maintain a blacklist of IP addresses that where collaborators in a recent DOS attack. A possible scenario, is one where an attacker wishes to sway victims to use a malicious WI-FI network instead of a legitimate WI-FI network that is in the area. In our scenario, the attacker leverages the anti DOS mechanism to achieve this by performing the following attack: (1) attacker connects a device to a legitimate WI-FI. (2) Attacker attempts to perform a DOS attack on a WebRTC application from the legitimate WI-FI network. (3) Legitimate WI-FI external IP is blocked by the server. (4) Attacker creates a rouge WI-FI network aimed to exploit anyone that connects to it. (5) Clients, being unable to connect to the WebRTC application due to the IP blocking, may decide to connect to the rouge network in order to use the WebRTC application.

**Possible outcome:** sensitive information theft or privilege elevation on victim's devices.

## 5. WebRTC IN TELECOMMUNICATION

As demonstrated in Section 1, WebRTC has gained popularity among many applications and is being deployed heavily. The growing trend has sparked many innovative products. The focus of this section is inspecting WebRTC as an underlying framework for telecommunication companies.

## 5.1 WebRTC-Tele-Model Security Concerns

Implementing a WebRTC gateway network architecture (as illustrated in Figure 2) where communications are funneled to and from telecommunication servers means that in practice, conversations are decrypted, and re-encrypted on these servers—turning them into valuable targets of attack. Implications may include the copying of information, as well as interchanging call participants without peers' consent or knowledge.

## 5.2 WebRTC-Tele-Model Efficiency Concerns

Assuming there are more peers than servers, two possible architectures are feasible: (1) communication between peers, ran by one server alone, and (2) communication between peers, ran by several company-owned servers before reaching a peer. The advantage of the second solution is in the fact that it allows bulking conversations together and compressing their volume. If performed correctly, this may lower bandwidth throttle and thus company costs. At this point, it is an open question as to what is the best approach to take in this case.

## 5.3 Lawful Interception for WebRTC via IP Multimedia Subsystem Using 3rd Party Gateway

Although there are numerous WebRTC communication systems architectures, it is inevitable that the WebRTC IP Multimedia Subsystem (IMS) will take over and telcos will have to implement their WebRTC solutions in this model [15].

IMS is a functional architecture for multimedia service delivery. It was design by the Third Generation Partnership Project (3GPP) and later on by TISPAN, the standardization body of European Telecommunications Standards Institute (ETSI). IMS allows easier network management, service implementation (billing, provision, provisioning etc.) and user management (Authentication, Authorization, and Accounting) [16]. Therefore implementing WebRTC as IMS architecture will be the most cost-effective solution for telcoes products [15].

WebRTC IMS standardization is an ongoing process since 2012 conducted by 3GPP [17] and by the ETSI [18], Still there are open issues considering the IMS implementation and the lawful interception (LI) customizations. Meanwhile the 3GPP and the ESTI design the IMS components for WebRTC as:

- WIC (WebRTC IMS Client)
- WWSF (WebRTC Web Server Function)
- WAF (WebRTC Authorization Function)
- eP-CSCF (P-CSCF enhanced for WebRTC)
- eIMS-AGW (IMS Access Gateway enhanced for WebRTC)

We assume that most of the WebRTC Communication Service Providers (CSP) will assist a 3rd party WebRTC Gateway Provider (GWP) to establish the connection with internal and external WebRTC clients [19], it's still not clear how the responsibility of WebRTC LI will be divided between the CSP and the 3rd party GWP.

IMS architecture solves many of the WebRTC standard requirements among them is lawful interception, which is already defined for IMS by the 3GPP and ETSI [20] and is globally accepted. Implementing LI solutions for different WebRTC architectures will be infeasible for CSPs due to the complexity the architecture and the lack of standardization of the WebRTC protocol.

The CSP should provide LI from the point when the commercial service is established and they shall co-operate immediately with the Law Enforcement Agency (LEA) and able a provision of targets to be on real-time basis. The ETSI also stated that the LI requirements are also relevant to 3rd party CSP, in our case the GWP.

As we see the WebRTC IMS architecture, the general presence of responsibilities and information between the CSP and the GWP should be as the following: the CSP will possess most of the information related the target's identity and enabled services where the GWP will be responsible to the delivery of the communication with other clients. Because WebRTC is established in RTCP most of the communication provision should be handled by the GWP, nevertheless it should be agreed in advance between the CSP, GWP and the LEA [20].

Under the ETSI current design for the WebRTC IMS [18], we propose the following (see Table 2) allocation of responsibilities between the CSP and the GWP (Figure 7) on data shared with the LEA.

**Table 2:** LI allocation of responsibilities - CSP and GWP

| Group | Item | Responsible |
|---|---|---|
| **General Intercepted information** | | |
| Specific identifiers for LI | Lawful interception identifier (LIID) – shall uniquely identify the target in the SP such as Telephone number or URI | CSP |
| | Network identifier | CSP |
| Location (changes between LEAs) | Current geographic | CSP |
| | Physical location | CSP |
| Intercepted information | Identities that attempted telecommunication with the target identity | GWP |
| | Identities used by the target | CSP |
| | Details of services used and their parameters | CSP |
| **Intercept related information (IRI)** | | |
| Events – IRI packet | target | CSP |
| | Second party | GWP |
| Correlation indications of IMS IRI | Call-ID, From tag, To tag | GWP |
| | IMS Charging ID | CSP |
| Correlation indications | Call-ID, From tag, To tag | GWP |
| | IMS Charging ID | CSP |
| | Connection addresses and ports | GWP |
| **Content of Communication (CC)** | | |
| RTP multimedia | RTP multimedia packet- RTP header, UDP header and IP header | GWP |
| | Copy of the multimedia stream also correlated with CIN | GWP |
| Encryption | Remove encryption applied on content | GWP |
| | Provide decryption keys that are available | GWP |
| **Security aspects** | | |
| General | build-in in the IMS architecture | CSP+GWP |

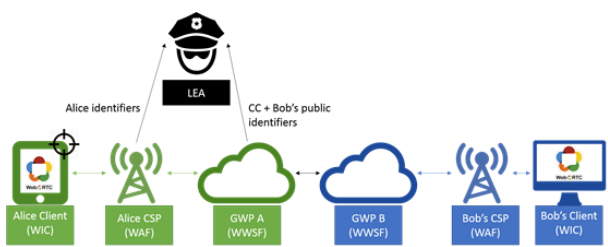

**Figure 7:** Alice uses WebRTC IMS via CSP and is under the local LEA provision.

## 6. RELATED WORKS

Rescorla (2013) reviewed the WebRTC main threats and formatted a technical security architecture models [21] [22] to deal with them. Rescorla's drafts map the current security state of WebRTC, suggesting specific technical solution for each threat. For example, to deal with communication security he recommended using DTLS/DTLS-SRTP and not allowing JavaScript to access to computations of the DTLS key. For Authentication, he detailed an IdP Mechanism, and for DDOS he suggested implementing a sensible flow control.

Based on Rescorla's drafts, Werner (2013) proposed security mechanisms which stand for the CIA information security principals [2]. The framework is built from four components: (1) Server– delivers the application and enables the signaling between the browsers (2) Browser – the runtime environment (3) Signaling path – http(s) or webSockets (4) media path – the data transformation protocol.

Mauro and Longo (2015) had successfully recognized encrypted WebRTC sessions using decision theory together with typical traffic features such as: inter-arrival times, packet lengths and the number of sent and received packets. In the future, we believe that Machine Learning can be used to extract information about the user's activity, the length of the conversation and the transferred objects (audio, video, files etc.) [14]

López-Fernández et al. (2014) investigated authorizations models in WebRTC. A Bad authorization model might financially harm a company (users can gain free access to paid services) and also damage the user's privacy (attacker could masquerade as the user) [23]. They explored two authorization models for WebRTC: (1) access control lists (ACLs), and (2) capability-based security (CAP). ACL control is that in which users rescue permission on an object and CAP uses a specific token for a permission. While CAP is faster, ACL has more functionality.

Koistinen (2014) explained and demonstrated how to implement the DTLS-SRTP functionalities to the traditional WebRTC browser to browser (P2P) model and measures the encryption impact on the performance [24]. The results shows that a single (S)RTP calls packet processing delay mostly under one millisecond time, Koistinen claims that this kind of delay is acceptable in the real world.

Johnston, *et al*. (2013) claimed that enterprise might have difficulties dealing with WebRTC due to restrictive enterprise firewall or untraceable P2P connection [25]. They suggest some general approaches to dealing with this issue but concluded that there is still missing enterprise authorization and application of WebRTC traffic policy.

Aghila and Chandirasekaran (2011) showed that SDES, ZRTP and MICKY are prone to Man-in-The-Middle attacks. Therefore WebRTC was designed with a default encryption of DTLS-SRTP. In this research we based our findings on the similar encryption problems of the SDES [26].

**Table 3:** Summary of related works.

| Reference | Title | Description |
|---|---|---|
| [21] | WebRTC Security Architecture | Technical security measures against various threats |
| [22] | Security Considerations for WebRTC | A general overview of WebRTC threats |
| [2] | WebRTC Security in the context of a DHT implementation | A general implementation of a security solution to deal with Rescorla Security Considerations |
| [14] | A Decision Theory Based Tool for Detection of Encrypted WebRTC Traffic | Prediction of WebRTC protocol sessions in the wild using Decision Theory. |
| [23] | Authentication, Authorization, and Accounting in WebRTC PaaS Infrastructures | Review of two authorizations models (CAP/ACL) |
| [24] | Implementation and Evaluation of Security on a Gateway for Web-based Real-Time Communication | Implementation and performance review on DTLS-SRTP functionalities |
| [25] | Taking on WebRTC in an Enterprise | Enterprises' challenges to dealing with WebRTC authorization and application of a security policy |
| [26] | An Analysis of VoIP Secure Key Exchange Protocols Internet | Penetration testing on SDES, ZRTP and MICKY shows that DTLS-SRTP is also immune against MITM attack. |

## 7. CONCLUSIONS

WebRTC is a good example of a properly designed technology; it shows a rapid deployment as well as high flexibility and adaptation to its surrounding protocols, encryption schemes and hosting environments. To most, WebRTC displays sharp abilities when it comes to privacy and security. But WebRTC is not a standalone solution; it requires a host application, a hosting server and a transportation layer connecting them. In this article we have demonstrated attacks against each of these underlying components which shows that while a WebRTC application comes with a set of strong security features, it is not immune to direct attack against its components, as well as its vulnerabilities on its underlying host application.